# Jerk and Hyperjerk in a Rotating Frame of Reference

**Amelia Carolina Sparavigna**
Department of Applied Science and Technology, Politecnico di Torino, Italy.

**Abstract:** Jerk is the derivative of acceleration with respect to time and then it is the third order derivative of the position vector. Hyperjerks are the n-th order derivatives with n>3. This paper describes the relations, for jerks and hyperjerks, between the quantities measured in an inertial frame of reference and those observed in a rotating frame. These relations can be interesting for teaching purposes.

**1. Introduction**
Jerk is the rate of change of acceleration, that is, it is the derivative of acceleration with respect to time [1,2]. In teaching physics, jerk is often completely neglected; however, as remarked in the Ref.1, it is a physical quantity having a great importance in practical engineering, because it is involved in mechanisms having rotating or sliding pieces, such as cams and genevas. For this reason, jerks are commonly found in models for the design of robotic arms. Moreover, these derivatives of acceleration are also considered in the planning of tracks and roads, in particular of the track transition curves [3]. Let us note that, besides in mechanics and transport engineering, jerks can be used in the study of several electromagnetic systems, as proposed in the Ref.4. Jerks of the geomagnetic field are investigated too, to understand the dynamics of the Earth's fluid, iron-rich outer core [5].
Jerks and hyperjerks - jerks having a n-th order derivative with n>3 - are also attractive for researchers that are studying the behaviour of complex systems. In fact, in 1997, Linz [6] and Sprott [7] generalized jerks into the "jerk functions", functions which are used to describe systems able of displaying a chaotic behaviour. In [8], it had been shown that the hyperjerk systems are prototypical examples of complex dynamical systems in a high-dimensional phase space.
The jerk is important for engineering because it is able evaluating the destructive effects of motion on a mechanism or the uneasiness feeling caused to the passengers in vehicles [9]. In fact, when designing a train, engineers consider to keep the jerk less than 2 metres per second cubed for the passengers' comfort. It is therefore important, in teaching physics for engineering students, to devote some time to the discussion of jerk. Here we are proposing, in particular, the discussion of the relations for jerks and hyperjerks, between the quantities measured in an inertial frame of reference and those observed in a rotating frame. This is a subject that, to the author's best knowledge, had received no attention from physics teachers.

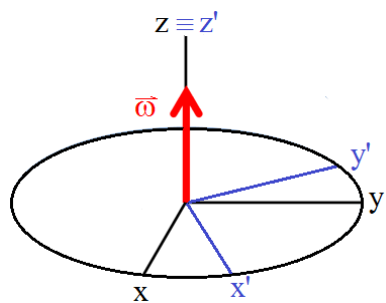

Figure 1: Cartesian inertial frame and rotating frame, having the same origin at rest.

## 2. Jerks and Hyperjerks

Let us consider two frames of reference having the same origin at rest. One is an inertial frame, $(x, y, z)$, the other a rotating frame $(x', y', z')$, which is rotating about $z=z'$ with constant angular velocity $\vec{\omega} = \omega \hat{k}$ (see Figure 1). $\hat{i}, \hat{j}, \hat{k}$ and $\hat{i}', \hat{j}', \hat{k}'$ are the unit vectors of axes. For any point in the space, we have:

$$\vec{r} = \vec{r}'$$
$$x\hat{i} + y\hat{j} + z\hat{k} = x'\hat{i}' + y'\hat{j}' + z'\hat{k}' \tag{1}$$

Let us derive (1):

$$\frac{d}{dt}(x\hat{i} + y\hat{j} + z\hat{k}) = \frac{d}{dt}(x'\hat{i}' + y'\hat{j}' + z'\hat{k}') \tag{2}$$

Accordingly, the velocity is:

$$\vec{v} = \frac{dx}{dt}\hat{i} + \frac{dy}{dt}\hat{j} + \frac{dz}{dt}\hat{k} = \frac{dx'}{dt}\hat{i}' + \frac{dy'}{dt}\hat{j}' + \frac{dz'}{dt}\hat{k}' + x'\frac{d\hat{i}'}{dt} + y'\frac{d\hat{j}'}{dt} + z'\frac{d\hat{k}'}{dt}$$

$$= \vec{v}' + \vec{\omega} \times \vec{r}' \tag{3}$$

Let us stress that the unit vectors $\hat{i}', \hat{j}', \hat{k}'$ are rotating. Therefore, in (3), we used the derivatives of unit vectors:

$$\frac{d\hat{i}'}{dt} = \vec{\omega} \times \hat{i}'; \quad \frac{d\hat{j}'}{dt} = \vec{\omega} \times \hat{j}'; \quad \frac{d\hat{k}'}{dt} = \vec{\omega} \times \hat{k}' \tag{4}$$

So we have the well-known equation between the velocities observed in the two frames:

$$\vec{v} = \vec{v}' + \vec{\omega} \times \vec{r}' \tag{5}$$

Let us derive this equation again:

$$\frac{d}{dt}(v_x \hat{i} + v_y \hat{j} + v_z \hat{k}) = \frac{d}{dt}(\vec{v}' + \vec{\omega} \times \vec{r}') \tag{6}$$

We can easily find:

$$\vec{a} = \vec{a}' + 2\vec{\omega} \times \vec{v}' + \vec{\omega} \times \vec{\omega} \times \vec{r}' \tag{7}$$

In (7), we have the Coriolis and centripetal terms. These are the accelerations we can create by means of the vectors at our disposal, which are $\vec{\omega}, \vec{v}'$ and $\vec{r}'$.

The jerk is the derivative of the acceleration and then:

$$\vec{J} = \frac{d}{dt}(a_x\hat{i} + a_y\hat{j} + a_z\hat{k}) = \frac{d}{dt}(\vec{a}' + 2\vec{\omega}\times\vec{v}' + \vec{\omega}\times\vec{\omega}\times\vec{r}')$$

$$= \frac{da'_x}{dt}\hat{i}' + \frac{da'_y}{dt}\hat{j}' + \frac{da'_z}{dt}\hat{k}' + a'_x\frac{d\hat{i}'}{dt} + a'_y\frac{d\hat{j}'}{dt} + a'_z\frac{d\hat{k}'}{dt} + 2\vec{\omega}\times\frac{d\vec{v}'}{dt} + \vec{\omega}\times\vec{\omega}\times\frac{d\vec{r}'}{dt}$$

(8)

To avoid any confusion with the unit vector of y-axis, here the symbol of the jerk is written using the upper-case letter. The first of terms referring to the rotating frame is the jerk seen in it.

$$\vec{J} = \vec{J}' + a'_x\frac{d\hat{i}'}{dt} + a'_y\frac{d\hat{j}'}{dt} + a'_z\frac{d\hat{k}'}{dt} + 2\vec{\omega}\times\frac{d\vec{v}'}{dt} + \vec{\omega}\times\vec{\omega}\times\frac{d\vec{r}'}{dt} \quad (9)$$

Using (4) again, we have the equation linking jerks in the two frames:

$$\vec{J} = \vec{J}' + \vec{\omega}\times\vec{a}' + 2\vec{\omega}\times\vec{a}' + 2\vec{\omega}\times\vec{\omega}\times\vec{v}' + \vec{\omega}\times\vec{\omega}\times\vec{v}' + \vec{\omega}\times\vec{\omega}\times\vec{\omega}\times\vec{r}'$$

(10)

$$= \vec{J}' + 3\vec{\omega}\times\vec{a}' + 3\vec{\omega}\times\vec{\omega}\times\vec{v}' + \vec{\omega}\times\vec{\omega}\times\vec{\omega}\times\vec{r}'$$

In (10), we have one term more, because, besides $\vec{\omega}, \vec{v}'$ and $\vec{r}'$, we have also $\vec{a}'$.
In the Figure 2 we can see the direction of vector $\vec{\omega}\times\vec{\omega}\times\vec{\omega}\times\vec{r}'$, for instance.

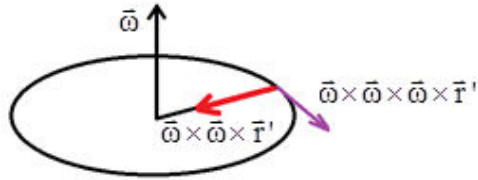

Figure 2: One of the terms in Eq.10.

According to [9], there is no universally accepted name for the fourth derivative, which is the rate of the change of jerk. The term "jounce" has been used, but this term has the drawback of using the same initial letter as jerk. Another suggestion is "snap" (symbol s). The "crackle" (symbol c) and "pop" (symbol p) are used for 5-th and 6-th derivatives respectively.

Let us determine the relation for the snaps. We have to derive again:

$$\frac{d\vec{J}'}{dt} = \frac{d}{dt}\left(\vec{J}' + 3\vec{\omega}\times\vec{a}' + 3\vec{\omega}\times\vec{\omega}\times\vec{v}' + \vec{\omega}\times\vec{\omega}\times\vec{\omega}\times\vec{r}'\right) \quad (11)$$

Then:

$$\vec{S} = \vec{S}' + \vec{\omega} \times \vec{J}' + 3\vec{\omega} \times \vec{J}' + 3\vec{\omega} \times \vec{\omega} \times \vec{a}' + 3\vec{\omega} \times \vec{\omega} \times \vec{a}' \\ + 3\vec{\omega} \times \vec{\omega} \times \vec{\omega} \times \vec{v}' + \vec{\omega} \times \vec{\omega} \times \vec{\omega} \times \vec{v}' + \vec{\omega} \times \vec{\omega} \times \vec{\omega} \times \vec{\omega} \times \vec{r}' \quad (12)$$

The equation between the snaps observed in the two frames is:

$$\vec{S} = \vec{S}' + 4\vec{\omega} \times \vec{J}' + 6\vec{\omega} \times \vec{\omega} \times \vec{a}' + 4\vec{\omega} \times \vec{\omega} \times \vec{\omega} \times \vec{v}' + \vec{\omega} \times \vec{\omega} \times \vec{\omega} \times \vec{\omega} \times \vec{r}' \quad (13)$$

In the Figure 3 we can see the direction of vector $\vec{\omega} \times \vec{\omega} \times \vec{\omega} \times \vec{\omega} \times \vec{r}'$.

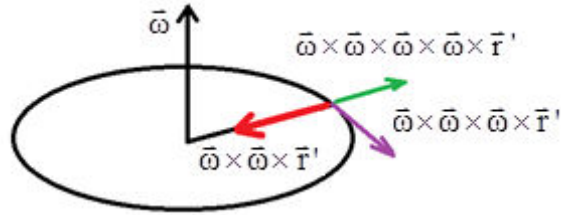

Figure 3: One of the terms in Eq.13.

We can easily continue for the following derivatives. For crackles, the relation is:

$$\vec{C} = \vec{C}' + 5\vec{\omega} \times \vec{S}' + 10\vec{\omega} \times \vec{\omega} \times \vec{J}' + 10\vec{\omega} \times \vec{\omega} \times \vec{\omega} \times \vec{a}' + 5\vec{\omega} \times \vec{\omega} \times \vec{\omega} \times \vec{\omega} \times \vec{v}' \\ + \vec{\omega} \times \vec{\omega} \times \vec{\omega} \times \vec{\omega} \times \vec{\omega} \times \vec{r}'$$

$$(14)$$

**3. Two simple examples**
Let us apply in two simple cases, the relations that we have previously discussed. The simplest possible examples are those of particles at rest. We can start from the inertial frame: let us consider a particle at rest on x-axis at distance R from the origin. We have that:

$$\vec{r} = \vec{r}' \Rightarrow R\,\hat{i} = R\cos(\omega t)\hat{i}' - R\sin(\omega t)\hat{j}' = RC_\omega\,\hat{i}' - RS_\omega\,\hat{j}' \quad (15)$$

In (15), we defined $C_\omega = \cos(\omega t), S_\omega = \sin(\omega t)$. We have velocity and acceleration:

$$\vec{v} = 0 \\ \vec{v}' = -\vec{\omega} \times \vec{r}' = -\omega\hat{k} \times \left(RC_\omega\,\hat{i}' - RS_\omega\,\hat{j}'\right) = -\omega RS_\omega\,\hat{i}' - \omega RC_\omega\,\hat{j}' \quad (16)$$

From the relation $\vec{a} = \vec{a}' + 2\vec{\omega} \times \vec{v}' + \vec{\omega} \times \vec{\omega} \times \vec{r}'$:

$$\vec{a} = 0 \\ \vec{a}' = -2\vec{\omega} \times \vec{v}' - \vec{\omega} \times \vec{\omega} \times \vec{r}' = 2\vec{\omega} \times \vec{\omega} \times \vec{r}' - \vec{\omega} \times \vec{\omega} \times \vec{r}' = -\omega^2 RC_\omega\,\hat{i}' + \omega^2 RS_\omega\,\hat{j}'$$

$$(17)$$

And, of course, we have from (10), a jerk:

$$\vec{J} = 0$$
$$\vec{J}' = -3\vec{\omega} \times \vec{a}' - 3\vec{\omega} \times \vec{\omega} \times \vec{v}' - \vec{\omega} \times \vec{\omega} \times \vec{\omega} \times \vec{r}'$$
$$= -3\vec{\omega} \times \vec{\omega} \times \vec{\omega} \times \vec{r}' + 3\vec{\omega} \times \vec{\omega} \times \vec{\omega} \times \vec{r}' - \vec{\omega} \times \vec{\omega} \times \vec{\omega} \times \vec{r}' = \omega^3 R S_\omega \hat{i}' + \omega^3 R C_\omega \hat{j}'$$

(18)

Jerk exists because the particle, in the rotating frame, has a centripetal acceleration, which is a vector having constant modulus but a direction changing with time. Also in the case of a particle in uniform motion or in uniformly accelerated motion, in the rotating frame we observe this particle having a jerk. Snap, crackle, pop and so on are different from zero.
In the same manner, we can consider a particle at rest in a rotating frame:

$$\vec{r} = \vec{r}' \Rightarrow R\cos(\omega t)\hat{i} + R\sin(\omega t)\hat{j} = R C_\omega \hat{i} + R S_\omega \hat{j} = R\hat{i}' \qquad (19)$$

We have velocity and acceleration:

$$\vec{v}' = 0$$
$$\vec{v} = \vec{\omega} \times \vec{r}' = \omega \hat{k} \times (R\hat{i}') = \omega R \hat{j}' \qquad (20)$$

$$\vec{a}' = 0$$
$$\vec{a} = \vec{\omega} \times \vec{\omega} \times \vec{r}' = \vec{\omega} \times (\omega R \hat{j}') = -\omega^2 R \hat{i}' \qquad (21)$$

In (20) and (21), we easily recognize the well-known velocity and acceleration of the circular uniform motion. However, we have also a jerk:

$$\vec{J}' = 0$$
$$\vec{J} = \vec{\omega} \times \vec{\omega} \times \vec{\omega} \times \vec{r}' = \vec{\omega} \times (-\omega^2 R \hat{i}') = -\omega^3 R \hat{j}' \qquad (22)$$

And crackle and pop:

$$\vec{C}' = 0; \quad \vec{C} = \vec{\omega} \times \vec{\omega} \times \vec{\omega} \times \vec{\omega} \times \vec{r}' = \vec{\omega} \times (-\omega^3 R \hat{j}') = \omega^4 R \hat{i}'$$
$$\vec{P}' = 0; \quad \vec{P} = \vec{\omega} \times \vec{\omega} \times \vec{\omega} \times \vec{\omega} \times \vec{\omega} \times \vec{r}' = \vec{\omega} \times (\omega^4 R \hat{i}') = \omega^5 R \hat{j}'$$

(23)

As we have seen in this example, jerk exists because the particle has a centripetal acceleration, which is a vector having constant modulus but a direction changing with time. We can propose, for instance, some problems on carousels or chair-o-planes, asking the students to evaluate the maximum value of angular speed to keep the jerk less than 2 metres per second cubed. Of course, these two proposed examples are quite simple: they can be used by students for practicing with rotating frames and cross products.